\documentstyle[seceq,epsf,preprint]{ptptex}

\newcommand{\bra}[1]{\langle {#1} |}     
\newcommand{\ket}[1]{| {#1} \rangle}     

\title{
A Possible Extension of a Trial State in the TDHF Theory with Canonical Form 
in the Lipkin Model}
\subtitle{
Canonicity conditions in an extended state of the $su(2)$-coherent state
}    

\author{
Yasuhiko {\sc Tsue}$^1$ and Hideaki {\sc Akaike}$^2$
}

\inst{
$^1$Physics Division, Faculty of Science, Kochi University, Kochi 780-8520, 
Japan\\
$^2$Department of Applied Science, Kochi University, Kochi 780-8520, 
Japan
}

\recdate{
\today
}

\abst{
With the aim of the extension of the TDHF theory in the canonical form 
in the Lipkin model, 
the trial state for the variation is constructed, which is an extension 
of the Slater determinant. The canonicity condition 
is imposed to formulate the variational approach in the canonical form. 
A possible solution of the canonicity condition is given and the zero-point 
fluctuation induced by the uncertainty principle is investigated in terms 
of the minimum uncertainty relation. 
As an application, the ground state energy is evaluated. 
}

\begin{document}

\maketitle

\section{Introduction}

The time-dependent Hartree-Fock (TDHF) theory is one of powerful methods to 
investigate the dynamics of quantum many-fermion systems. 
Especially, this theory has been developed in nuclear many-body 
problems.\cite{RS80}
The TDHF theory and the Hartree-Fock approximation are formulated 
based on the variational method. 
In these methods, the trial 
state for the variation is prepared to describe the many-fermion system. 
The Slater determinant is usually adopted as a possible trial state. 
This state gives a possible classical counterpart of the quantum many-fermion 
system. For this purpose, the TDHF theory with the canonical form presents 
a suitable treatment. In this treatment, the canonicity condition 
plays an essential and a central role.\cite{YK87,MMSK80}
This trial state however 
may be regarded as a kind of the coherent state. 

On the other hand, 
in the many-boson systems, the coherent state also gives the classical 
image of quantum many-boson systems. We have been formulated the 
time-dependent variational approach to quantum many-boson systems 
including appropriate quantum fluctuations for the systems under 
consideration.\cite{TFKY91,KPTY01}
Then, the squeezed state is applied to the variation as a possible 
trial state. 

In quantum many-fermion systems, one of the present authors (Y.T.) together 
with Yamamura and Kuriyama have 
constructed the trial state in both 
the pairing\cite{TKY94} and the Lipkin models\cite{TAKY96} 
corresponding to the boson squeezed state. In these models, the Hamiltonian 
can be expressed in terms of the quasi-spin operators. 
Then, the Slater determinantal state is identical with the $su(2)$-coherent 
state. 
In this sense, we thus call the extended trial state the quasi-spin 
squeezed state. 
We have constructed the variational approximation, which include the result 
of the Hartree-Fock approximation, in the Lipkin model.\cite{TAKY96}
Then, this variational method using the quasi-spin squeezed state gives 
the results obtained in the random phase approximation (RPA) in the 
certain approximation.\cite{NASA,KPTY01} 
As a result, our quasi-spin squeezed state approach to the Lipkin model 
is a possible extension of the Hartree-Fock approximation. 

In this paper, with the aim of extension of our previous work 
to the time-dependent variational approach to the Lipkin model, 
we investigate a possible solution of the canonicity condition 
in the quasi-spin squeezed state. The canonicity condition plays a central 
role to formulate the TDHF theory in the canonical form. 
Thus, we can construct the extended TDHF theory in the canonical form, 
if we use the 
quasi-spin squeezed state as a trial state instead of the Slater determinant. 
Also, the effect of the zero-point oscillation induced by the uncertainty 
principle is investigated in terms of the canonical variables. 
In this paper, 
the ground state energy is calculated by imposing a condition 
of minimum uncertainty relation in order to consider 
the above-mentioned zero-point oscillation. 
The comparison of the ground state energies obtained by various states, 
except for the quasi-spin squeezed state investigated in this paper, 
is also reported in \citen{YKPPT03}. 

This paper is organized as follows. 
In the next section, the Lipkin model is recapitulated 
containing the notations. 
In \S 3, the Slater determinant is used to describe the Lipkin model. 
In \S 4, a possible extension of the trial state for the variation 
is given. This state corresponds to the boson coherent state in the 
many-boson systems. Further, the canonicity conditions are imposed and 
a possible solutions of these conditions are given. Also, the way to 
obtain the approximate solution is discussed. 
The original idea to solve the canonicity condition is found in Ref.\citen{Y}. 
In \S 5, the energy expectation value is calculated including 
the zero-point fluctuation induced by the uncertainty principle. 
The last section is devoted to a summary.

\section{Recapitulation of the Lipkin model}

In this paper, we give a possible extension of the TDHF theory 
in the case of the Lipkin model. We consider $2\Omega$ fermions moving 
in two single-particle levels with the same degeneracy $2\Omega$. 
Here, $\Omega$ is a positive integer, and for the convenience of later 
treatment, we use a half-integer $j$ defined by 
$\Omega=j+1/2$ and an additional quantum number $m$ to distinguish each 
single-particle state. 
As the free vacuum $\ket{0}$, we can adopt a state in which one level is 
occupied by all fermions under consideration. 
This level may be called hole-level and the other particle-level. 
In this model, we introduce the following set of operators : 
\begin{eqnarray}\label{2-1}
& &{\hat S}_+=\sum_m a_{jm}^* (-)^{j-m}b_{j-m}^* \ , \nonumber\\
& &{\hat S}_-=\sum_m (-)^{j-m}b_{j-m} a_{jm} \ , \nonumber\\
& &{\hat S}_0=1/2\cdot\sum_m (a_{jm}^* a_{jm}+b_{jm}^* b_{jm})
-\Omega \ . 
\end{eqnarray}
Here, $m$ runs from $-j$ to $+j$ and $(a_{jm}^* , a_{jm})$ and 
$(b_{jm}^* , b_{jm})$ denote particle and hole operators in the 
particle and hole state $jm$, respectively. 
They are fermion operators. The set $({\hat S}_+, {\hat S}_- , {\hat S}_0)$ 
satisfies the $su(2)$ algebra obeying the relations 
\begin{equation}\label{2-2}
[\ {\hat S}_- \ , \ {\hat S}_+\ ]=-2{\hat S}_0 \ , \qquad
[\ {\hat S}_0 \ , \ {\hat S}_{\pm}\ ]=\pm{\hat S}_{\pm} \ . 
\end{equation}

In order to give a transparent connection to boson system, 
which we have already given the form, it may be convenient to define 
the quantities 
\begin{equation}\label{2-3}
{\hat A}^*={\hat S}_+/\sqrt{2\Omega} \ , \quad
{\hat A}={\hat S}_-/\sqrt{2\Omega} \ , \quad
{\hat N}=2(\Omega+{\hat S}_0) \ . 
\end{equation}
The set $({\hat A}^*, {\hat A}, {\hat N})$ satisfies the relations 
\begin{equation}\label{2-4}
[\ {\hat A} \ , \ {\hat A}^* \ ]=1-{\hat N}/2\Omega \ , \quad
[\ {\hat N} \ , \ {\hat A}^*\ ]=+2{\hat A}^* \ , \quad 
[\ {\hat N} \ , \ {\hat A}\ ]=-2{\hat A} \ . 
\end{equation}
The first relation shows that if ${\hat N}/2\Omega$ is negligible, the 
operators ${\hat A}$ and ${\hat A}^*$ can be regarded as boson operators. 
Further, we define ${\hat Q}$, ${\hat P}$ and ${\hat R}$ in the 
following forms : 
\begin{eqnarray}\label{2-5}
{\hat Q}&=&\sqrt{\hbar/2}\cdot ({\hat A}^*+{\hat A}) \nonumber\\
&=&\sqrt{\hbar/2}\cdot ({\hat S}_++{\hat S}_-)/\sqrt{2\Omega}
=\sqrt{\hbar/\Omega}\cdot{\hat S}_x \ , \nonumber\\
{\hat P}&=&i\sqrt{\hbar/2}\cdot ({\hat A}^*-{\hat A}) \nonumber\\
&=&i\sqrt{\hbar/2}\cdot ({\hat S}_+-{\hat S}_-)/\sqrt{2\Omega}
=-\sqrt{\hbar/\Omega}\cdot{\hat S}_y \ , \nonumber\\
{\hat R}&=&1-{\hat N}/2\Omega=-{\hat S}_0/\Omega=-{\hat S}_z/\Omega \ .
\end{eqnarray}
The operators ${\hat Q}$, ${\hat P}$ and ${\hat R}$ satisfy the relations 
\begin{eqnarray}
& &{\hat Q}^*={\hat Q} \ , \qquad {\hat P}^*={\hat P}\ , 
\qquad {\hat R}^*={\hat R} \ , 
\label{2-6}\\
& &[\ {\hat Q} \ , \ {\hat P} \ ]=i\hbar{\hat R} \ . 
\label{2-7}
\end{eqnarray}
In this case, also, if ${\hat N}/2\Omega$ is negligible, ${\hat Q}$ 
and ${\hat P}$ can be regarded as the coordinate and its canonical momentum 
and ${\hat R}$ becomes unit operator. 

For the operators ${\hat Q}$ and ${\hat P}$ satisfying the relations 
(\ref{2-6}) and (\ref{2-7}), we have the following uncertainty relation : 
\begin{equation}\label{2-8}
\Delta Q\cdot \Delta P \geq \hbar/2\cdot |\langle {\hat R} \rangle| \ . 
\end{equation}
Here, $\Delta Q$ and $\Delta P$ are defined by 
\begin{equation}\label{2-9}
\Delta Q=\sqrt{\langle ({\hat Q}-\langle {\hat Q} \rangle)^2\rangle} \ , 
\qquad
\Delta P=\sqrt{\langle ({\hat P}-\langle {\hat P} \rangle)^2\rangle} \ . 
\end{equation}
The symbol $\langle {\hat O} \rangle$ denotes the expectation value 
of the operator ${\hat O}$ for an arbitrary state $\ket{\ }$. 
This relation can be proved by preparing the relations 
\begin{equation}\label{2-10}
\langle {\hat Y}^*{\hat Y}\rangle \geq 0 \ , \quad ({\rm positive\ definite})
\end{equation}
where, for arbitrary real number $y$, ${\hat Y}^*$ and ${\hat Y}$ are 
defined as 
\begin{eqnarray}\label{2-11}
& &{\hat Y}^*=({\hat P}-\langle {\hat P} \rangle )y
+i({\hat Q}-\langle {\hat Q} \rangle) \ , \nonumber\\
& &{\hat Y}=({\hat P}-\langle {\hat P} \rangle )y
-i({\hat Q}-\langle {\hat Q} \rangle) \ .
\end{eqnarray}

\section{Slater determinant as a trial state for the variation}

First, we introduce the following state : 
\begin{equation}\label{3-1}
\ket{\phi(A)}=1/\sqrt{\Phi(A^*A)}\cdot \exp(A{\hat A}^*)\ket{0} \ , 
\end{equation}
where $\Phi(A^*A)$ is given by 
\begin{equation}\label{3-2}
\Phi(A^*A)=\bra{0}\exp(A^*{\hat A})\cdot \exp(A{\hat A}^*)\ket{0}
=(1+A^*A/2\Omega)^{2\Omega} \ . 
\end{equation}
The state $\ket{\phi(A)}$ is a Slater determinant with the condition 
\begin{equation}\label{3-3}
\bra{\phi(A)}\phi(A)\rangle=1\ . 
\end{equation}
The factor $\sqrt{\Phi(A^*A)}$ reduces to $\exp(A^*A/2)$ at the 
limit $A^*A/2\Omega \rightarrow 0$ and the state $\ket{\phi(A)}$ 
becomes a coherent state in boson system. 

With the help of the following canonicity condition, we introduce 
a set of canonical variables $(X^*, X)$ : 
\begin{equation}\label{3-4}
\bra{\phi(A)}\partial_X \ket{\phi(A)}=X^*/2 \ . 
\end{equation}
Of course, the variables $X^*$ and $X$ obey the Poisson bracket relation 
$\{ X, X^* \}_P=1$. 
Further, the equation of motion for $X^*$ and $X$ are given 
by the variational principle. 
The explicit calculation of the left-hand side of Eq.(\ref{3-4}) gives 
\begin{equation}\label{3-5}
\bra{\phi(A)}\partial_X \ket{\phi(A)}=1/(1+A^*A/2\Omega)\cdot
(A^*\partial_X A-A\partial_X A^*)/2 \ . 
\end{equation}
A possible solution of Eqs.(\ref{3-4}) and (\ref{3-5}) is given by 
\begin{equation}\label{3-6}
A^*=X^*/\sqrt{1-X^*X/2\Omega} \ , \qquad
A=X/\sqrt{1-X^*X/2\Omega} \ .
\end{equation}
With the use of the above relations (\ref{3-6}), we can express 
all relations in our present treatment in terms of $X^*$ and $X$.

The TDHF theory in the Lipkin model consists of the 
expectation values of the operators ${\hat A}^*$, ${\hat A}$ 
and ${\hat N}$ for the state $\ket{\phi(A)}$ : 
\begin{eqnarray}
& &\bra{\phi}{\hat A}^*\ket{\phi}=X^*\sqrt{1-X^*X/2\Omega} \ , 
\nonumber\\
& &\bra{\phi}{\hat A}\ket{\phi}=X\sqrt{1-X^*X/2\Omega} \ , 
\label{3-7}\\
& &\bra{\phi}{\hat N}\ket{\phi}=2X^*X \ , \qquad
\bra{\phi}1-{\hat N}/2\Omega\ket{\phi}=1-X^*X/\Omega \ . 
\label{3-8}
\end{eqnarray}
The above expectation values are for the state $\ket{\phi(A)}$. 
We can see that they are classical counterparts of the Holstein-Primakoff 
type boson representation of the $su(2)$ algebra. 
If $X^*X/2\Omega$ is negligible, the relations (\ref{3-7}) show that the 
expectation values of ${\hat A}^*$ and ${\hat A}$ are reduced to the canonical 
variables $X^*$ and $X$. 
Then, the factor $\sqrt{1-X^*X/2\Omega}$ can be attributed to the 
blocking effect, a kind of quantum effects, which comes from the exclusion 
principle. 
As was mentioned in \S 1, our main interest is to clarify the effect 
of the zero-point oscillation induced by the uncertainty principle. 
Therefore, as our terminology, we include the blocking effect 
in the classical counterpart.

With the aim of investigating the effect of the uncertainty principle, we calculate the square order of the expectation values of the operators 
${\hat A}^*$, ${\hat A}$ and ${\hat N}$ : 
\begin{eqnarray}
& &\bra{\phi}{\hat A}^{*2}\ket{\phi}
=X^{*2}(1-X^*X/2\Omega)(1-1/2\Omega) \ , \nonumber\\
& &\bra{\phi}{\hat A}^{2}\ket{\phi}
=X^{2}(1-X^*X/2\Omega)(1-1/2\Omega) \ , 
\label{3-9}\\
& &\bra{\phi}{\hat A}^{*}{\hat A}\ket{\phi}
=X^{*}X(1-X^*X/2\Omega)+(X^*X/2\Omega)^2 \ , \nonumber\\
& &\bra{\phi}{\hat A}{\hat A}^{*}\ket{\phi}
=X^{*}X(1-X^*X/2\Omega)+(1-X^*X/2\Omega)^2 \nonumber\\
& &\qquad\qquad\ \ 
=\bra{\phi}{\hat A}^*{\hat A}\ket{\phi}+(1-X^*X/\Omega) \ , 
\label{3-10}\\
& &\bra{\phi}(1-{\hat N}/2\Omega)^2\ket{\phi}
=(1-X^{*}X/\Omega)^2+1/\Omega\cdot X^*X/\Omega\cdot(1-X^*X/2\Omega)\ . \qquad
\label{3-11}
\end{eqnarray}
Using the above relations, we can set the following result for the state 
$\ket{\phi(A)}$ : 
\begin{eqnarray}
& &(\Delta Q)^2=\bra{\phi}{\hat Q}^2\ket{\phi}-\bra{\phi}{\hat Q}\ket{\phi}^2
\nonumber\\
& &\qquad\quad
=\hbar/2\cdot [1-(X^*X+X^{*2})/2\Omega][1-(X^*X+X^2)/2\Omega] \ , 
\label{3-12}\\
& &(\Delta P)^2=\bra{\phi}{\hat P}^2\ket{\phi}-\bra{\phi}{\hat P}\ket{\phi}^2
\nonumber\\
& &\qquad\quad
=\hbar/2\cdot [1-(X^*X-X^{*2})/2\Omega][1-(X^*X-X^2)/2\Omega] \ , 
\label{3-13}\\
& &\bra{\phi}{\hat Q}\ket{\phi}=\sqrt{\hbar/2}\cdot (X^*+X)
\sqrt{1-X^*X/2\Omega} \ , 
\label{3-14}\\
& &\bra{\phi}{\hat P}\ket{\phi}=i\sqrt{\hbar/2}\cdot (X^*-X)
\sqrt{1-X^*X/2\Omega} \ . 
\label{3-15}
\end{eqnarray}
With the use of the relations (\ref{2-5}), (\ref{3-8}), (\ref{3-12}) 
and (\ref{3-13}), we have the uncertainty relation 
\begin{eqnarray}\label{3-16}
(\Delta Q \cdot\Delta P)^2-(\hbar\langle {\hat R} \rangle/2)^2
&=&(\hbar/2)^2\cdot(1-X^*X/2\Omega)^2 \nonumber\\
& &\times [i(X^{*2}-X^2)/2\Omega]^2 \geq 0 \ .
\end{eqnarray}
From the above relation, we see that at the limit $X^*X/2\Omega \rightarrow 0$ 
or 1, $(\Delta Q)^2\rightarrow \hbar/2$, $(\Delta P)^2\rightarrow \hbar/2$ and 
$(\Delta Q \cdot \Delta P)^2-(\hbar\langle {\hat R}\rangle/2)^2
\rightarrow 0$. 
The above relations show us that the Slater determinant (\ref{3-1}) has 
the properties similar to those of the coherent state of the boson system 
with respect to the minimum uncertainty. 
Further, from Eqs.(\ref{3-12}) and (\ref{3-13}), we have 
\begin{eqnarray}
\langle({\hat P}^2+{\hat Q}^2)/2\rangle 
&=& (P^2+Q^2)/2 \nonumber\\
& &+\hbar/2\cdot[(1-X^*X/2\Omega)^2+(X^*X/2\Omega)^2] \ , 
\label{3-17}\\
(P^2+Q^2)/2&=&\hbar X^*X(1-X^*X/2\Omega) \ . 
\label{3-18}
\end{eqnarray}
Here, $Q$ and $P$ denote $\bra{\phi}{\hat Q}\ket{\phi}$ and 
$\bra{\phi}{\hat P}\ket{\phi}$, respectively. 
We can see that if $X^*X/2\Omega\rightarrow 0$ or 1, the above results 
are reduced to those given in the coherent state of the boson system. 
The parts $Q$ and $P$ show the classical parts, and the additional denote the 
effects of the zero-point oscillation.

Finally, we show the square of the quasi-spin, the results of which are 
\begin{eqnarray}
& &\langle {\hat S}_x \rangle^2 + \langle {\hat S}_y \rangle^2 + 
\langle {\hat S}_z \rangle^2 = \Omega^2 \ , 
\label{3-19}\\
& &\langle {\hat S}_x^2 \rangle + \langle {\hat S}_y^2 \rangle + 
\langle {\hat S}_z^2 \rangle = \Omega(\Omega+1) \ . 
\label{3-20}
\end{eqnarray}
Certainly, the above results show that the quantum fluctuations can be taken 
into account in our treatment. 
The above is given in the framework of the Slater determinant and it may be 
possible to give an understanding that the Slater determinant plays the same 
role as that of the coherent state in the boson system. 
Therefore, it cannot give a zero-point energy appropriate for the Hamiltonian, 
for example, 
\begin{equation}\label{3-21}
{\hat H}=\epsilon\cdot{\hat S}_0-\chi/4\Omega\cdot({\hat S}_+^2+{\hat S}_-^2)
\ .
\end{equation}
In order to give the appropriate zero-point energy, in the next section, 
we develop a possible extension of the TDHF theory.

\section{An extension of the trial state for the variation}

We extend the Slater determinant shown in Eq.(\ref{3-1}) to the form, 
which enable us to give a correct zero-point energy. 
The original idea was given by Yamamura in Ref.\citen{Y}. 
For this purpose, following the form shown in the boson system, 
we adopt the form 
\begin{equation}\label{4-1}
\ket{\psi(A,B)}=1/\sqrt{\Psi(B^*B)}\cdot\exp(B{\hat B}^{*2}/2)\ket{\phi(A)} 
\ . 
\end{equation}
Here, the operator ${\hat B}$ satisfies the condition 
\begin{equation}\label{4-2}
{\hat B}\ket{\phi(A)}=0 \ . 
\end{equation}
The normalization factor $\Psi(B^*B)$ is given as 
\begin{equation}\label{4-3}
\Psi(B^*B)=\bra{\phi(A)}\exp(B^*{\hat B}^{2}/2)\cdot\exp(B{\hat B}^{*2}/2)
\ket{\phi(A)} \ . 
\end{equation} 
Let us show possible forms of the operators ${\hat B}^*$ and ${\hat B}$ 
together with the factor $\Psi(B^*B)$. 
For this purpose, we introduce the following set of the operators 
\begin{eqnarray}\label{4-4}
& &{\hat T}_+=\sum_{m}\alpha_{jm}^*(-)^{j-m}\beta_{j-m}^* \ , 
\nonumber\\
& &{\hat T}_-=\sum_{m}(-)^{j-m}\beta_{j-m}\alpha_{jm} \ , 
\nonumber\\
& &{\hat T}_0=1/2\cdot\sum_{m}(\alpha_{jm}^*\alpha_{jm} + 
\beta_{jm}^* \beta_{jm})-\Omega \ . 
\end{eqnarray}
Here, $(\alpha_{jm}^*,\alpha_{jm})$ and $(\beta_{jm}^* , \beta_{jm})$ denote 
fermion operators. The vacuum is $\ket{\phi(A)}$. The explicit forms 
are as follows : 
\begin{eqnarray}\label{4-5}
& &\alpha_{jm}=Ua_{jm}-V(-)^{j-m}b_{j-m}^* \ , 
\quad (\alpha_{jm}\ket{\phi(A)}=0) \nonumber\\
& &\beta_{jm}=Ub_{jm}-V(-)^{j-m}a_{j-m}^* \ . 
\quad (\beta_{jm}\ket{\phi(A)}=0) 
\end{eqnarray}
Here, $U$ and $V$ are defined by 
\begin{equation}\label{4-6}
U=1/\sqrt{1+A^*A/2\Omega} \ , \qquad
V=A/\sqrt{2\Omega}\cdot 1/\sqrt{1+A^*A/2\Omega} \ . 
\end{equation}
They satisfy the relation $U^2+V^*V=1$. 
In the same way as that of the case (\ref{2-3}), we define the operator 
${\hat B}^*$ and ${\hat B}$ satisfying the relation (\ref{4-2}), 
together with ${\hat M}$, as 
\begin{eqnarray}\label{4-7}
& &{\hat B}^*={\hat T}_+/\sqrt{2\Omega} \ , \qquad
{\hat B}={\hat T}_-/\sqrt{2\Omega} \ , \nonumber\\
& &{\hat M}=2(\Omega+{\hat T}_0) \ . 
\end{eqnarray}
Clearly, we have ${\hat B}\ket{\phi(A)}=0$, and further, 
${\hat M}\ket{\phi(A)}=0$. 
They satisfy the algebra of the $su(2)$ and the commutation relations 
are given by 
\begin{equation}\label{4-8}
[\ {\hat B}\ , \ {\hat B}^*\ ]=1-{\hat M}/2\Omega\ , \quad
[\ {\hat M}\ , \ {\hat B}^*\ ]=+2{\hat B}^* \ , \quad
[\ {\hat M}\ , \ {\hat B}\ ]=-2{\hat B} \ . 
\end{equation}
With the use of the relations (\ref{4-8}), we can calculate 
$\Psi(B^*B)$ : 
\begin{equation}\label{4-9}
\Psi(B^*B)=1+\sum_{n=1}^{2\Omega}(2n-1)!!/(2^n\cdot n!)\cdot
\prod_{k=1}^{2n-1}(1-k/2\Omega)(B^*B)^n \ .
\end{equation}
If $\Omega\rightarrow \infty$, then $\Psi(B^*B)\rightarrow (1-B^*B)^{-1/2}$, 
which coincides with the case of the boson system. 
It should be noted that in the framework of the condition (\ref{4-2}), 
there are infinite possibilities for the selection of the operator ${\hat B}$. 
We adopt the form (\ref{4-7}) as one of the possibilities.

First, we introduce two sets of canonical variables $(X^*,X)$ and $(Y^*,Y)$ 
which satisfy the relations $\{X,X^*\}_P=\{Y,Y^*\}_P=1$ and 
$\{$ the other combinations $\}_P=0$. 
These variables obey the following canonicity conditions : 
\begin{eqnarray}
& &\bra{\psi(A,B)}\partial_X \ket{\psi(A,B)}=X^*/2\ , 
\label{4-10}\\
& &\bra{\psi(A,B)}\partial_Y \ket{\psi(A,B)}=Y^*/2\ . 
\label{4-11}
\end{eqnarray}
The left-hand side of the above relation is given by 
\begin{eqnarray}\label{4-12}
\bra{\psi(A,B)}\partial_Z\ket{\psi(A,B)}
&=&\Psi'(B^*B)/\Psi(B^*B)\cdot (B^*\partial_Z B-B\partial_Z B^*)/2 
\nonumber\\
& &+(1-2B^*B/\Omega\cdot \Psi'(B^*B)/\Psi(B^*B)) \nonumber\\
& &\times 1/(1+A^*A/2\Omega)\cdot (A^*\partial_Z A-A\partial_Z A^*)/2 \ . 
\end{eqnarray}
Here, $\Psi'(B^*B)$ denotes the derivative of $\Psi(B^*B)$ for $B^*B$ and 
$Z$ represents $X$ or $Y$. 
In order to get a possible solution of Eqs.(\ref{4-10}) and (\ref{4-11}), 
we set up the following equation : 
\begin{equation}\label{4-13}
\Psi'(B^*B)/\Psi(B^*B)\cdot (B^*\partial_Y B-B\partial_Y B^*)/2
=Y^*/2 \ .
\end{equation}
Then, $A^*$ and $A$ should obey 
\begin{equation}\label{4-14}
A^*\partial_Y A-A\partial_Y A^*=0 \ . 
\end{equation}
A solution of (\ref{4-13}) is obtained in the form 
\begin{equation}\label{4-15}
B^*=Y^*/\sqrt{K(Y^*Y)} \ , \qquad B=Y/\sqrt{K(Y^*Y)} \ , 
\end{equation}
where $K(Y^*Y)$ is given as a solution of the equation 
\begin{equation}\label{4-16}
K\Psi(Y^*Y/K)=\Psi'(Y^*Y/K) \ . 
\end{equation}
Then, we have
\begin{equation}\label{4-17}
1-2B^*B/\Omega\cdot \Psi'(B^*B)/\Psi(B^*B)=1-2Y^*Y/\Omega \ . 
\end{equation}
Since $B^*$ and $B$ are functions of only $Y^*$ and $Y$, the parameters 
$A^*$ and $A$ should satisfy the relation 
\begin{equation}\label{4-18}
(1-2Y^*Y/\Omega)\cdot 1/(1+A^*A/2\Omega)\cdot (A^*\partial_X A
-A\partial_X A^*)/2=X^*/2 \ .
\end{equation}
The above relation comes from Eq.(\ref{4-12}) for $Z=X$. 
A solution of Eq.(\ref{4-18}) is given by 
\begin{eqnarray}\label{4-19}
& &A^*=X^*/\sqrt{1-X^*X/2\Omega-4Y^*Y/2\Omega} \ , \nonumber\\
& &A=X/\sqrt{1-X^*X/2\Omega-4Y^*Y/2\Omega} \ .
\end{eqnarray}
The solution (\ref{4-19}) leads to the relation (\ref{4-14}). Thus, 
we can get the solution of the canonicity conditions (\ref{4-10}) and 
(\ref{4-11}) in the form (\ref{4-15}) and (\ref{4-19}). 
We can see that in the case of $Y^*=Y=0$, the results (\ref{4-19}) reduce 
to the forms (\ref{3-6}). 
With the use of the solutions (\ref{4-15}) and (\ref{4-19}), $U$ and $V$ 
defined in Eqs.(\ref{4-6}) are expressed as 
\begin{eqnarray}\label{4-20}
& &U=\sqrt{1-X^*X/2\Omega-4Y^*Y/2\Omega}\cdot 1/\sqrt{1-4Y^*Y/2\Omega} \ , 
\nonumber\\
& &V=X/\sqrt{2\Omega}\cdot 1/\sqrt{1-4Y^*Y/2\Omega} \ . 
\end{eqnarray}
Of course, $U$ and $V$ in Eqs.(\ref{4-20}) satisfy the relation 
$U^2+V^*V=1$.

We are now at the position to calculate the expectation values 
of the operators ${\hat A}^*$ and so on for the state $\ket{\psi(A,B)}$. 
For this purpose, first, we list up the relations 
\begin{eqnarray}
& &{\hat A}^*=\sqrt{2\Omega}UV^*(1-{\hat M}/2\Omega)
+U^2{\hat B}^*-V^{*2}{\hat B} \ , \label{4-21}\\
& &{\hat A}=\sqrt{2\Omega}UV(1-{\hat M}/2\Omega)
-V^2{\hat B}^*+U^{2}{\hat B} \ , \label{4-22}\\
& &{\hat N}=4\Omega V^*V(1-{\hat M}/2\Omega)
+2\sqrt{2\Omega}U(V{\hat B}^*+V^{*}{\hat B})+{\hat M} \ . 
\label{4-23}
\end{eqnarray}
Next, we show the expectation values of ${\hat B}^*$ and so on :
\begin{eqnarray}
& &\bra{\psi}{\hat B}^*\ket{\psi}
=\bra{\psi}{\hat B}\ket{\psi}=0 \ , 
\label{4-24}\\
& &\bra{\psi}{\hat M}\ket{\psi}=4Y^*Y \ , \quad
\bra{\psi}(1-{\hat M}/2\Omega)\ket{\psi}=1-4Y^*Y/2\Omega \ , 
\label{4-25}\\
& &\bra{\psi}{\hat B}^*{\hat B}\ket{\psi}=2(1-1/2\Omega)Y^*Y
-2/\Omega\cdot (Y^*Y)^2L(Y^*Y) \ , 
\label{4-26}\\
& &\bra{\psi}{\hat B}{\hat B}^*\ket{\psi}
=\bra{\psi}{\hat B}^*{\hat B}\ket{\psi}+(1-4Y^*Y/2\Omega) \ , 
\label{4-27}\\
& &\bra{\psi}{\hat B}^{*2}\ket{\psi}
=2Y^*\sqrt{K(Y^*Y)} \ , \qquad
\bra{\psi}{\hat B}^2\ket{\psi}
=2Y\sqrt{K(Y^*Y)} \ , 
\label{4-28}\\
& &\bra{\psi}{\hat M}^2\ket{\psi}
=16Y^*Y(1+Y^*Y\cdot L(Y^*Y)) \ , \nonumber\\
& &\bra{\psi}(1-{\hat M}/2\Omega)^2\ket{\psi}
=1-4/\Omega\cdot (1-1/\Omega)Y^*Y
+4/\Omega^2\cdot (Y^*Y)^2L(Y^*Y) \ . \qquad\ \ \ 
\label{4-29}
\end{eqnarray}
Here, $\bra{\psi}{\hat O}\ket{\psi}$ denotes the expectation value for the
state $\ket{\psi(A,B)}$. 
With the use of the derivative of $\Psi$ for $B^*B$, $L(Y^*Y)$ is 
defined as 
\begin{equation}\label{4-30}
K(Y^*Y)^2\cdot L(Y^*Y)=\Psi''(B^*B)/\Psi(B^*B) \ . 
\end{equation}
The simplest approximate forms of $K$ and $L$ are as follows : 
\begin{eqnarray}
& &K(Y^*Y)=(1-1/2\Omega)/2+(1-7/2\Omega+9/4\Omega^2)(Y^*Y)+\cdots \ ,
\label{4-31}\\
& &L(Y^*Y)=3(1-2/2\Omega)(1-3/2\Omega)/(1-1/2\Omega)\nonumber\\
& &\qquad\qquad
-24/\Omega\cdot(1-2/2\Omega)(1-3/2\Omega)(1-4/2\Omega)/(1-1/2\Omega)^2
\cdot(Y^*Y)+\cdots \ . \nonumber\\
& &\label{4-32}
\end{eqnarray}

With the use of the relations (\ref{4-20})$\sim$(\ref{4-25}), we can show the 
expectation values of ${\hat A}^*$, ${\hat A}$ and ${\hat N}$ : 
\begin{eqnarray}\label{4-33}
& &\bra{\psi}{\hat A}^*\ket{\psi}
=X^*\sqrt{1-X^*X/2\Omega-4Y^*Y/2\Omega} \ , \nonumber\\
& &\bra{\psi}{\hat A}\ket{\psi}
=X\sqrt{1-X^*X/2\Omega-4Y^*Y/2\Omega} \ , \nonumber
\\
& &\bra{\psi}{\hat N}\ket{\psi}
=2X^*X+4Y^*Y \ , \nonumber\\
& &\bra{\psi}1-{\hat N}/2\Omega\ket{\psi}
=1-X^*X/\Omega-2Y^*Y/\Omega \ . 
\end{eqnarray}
In addition to the above cases, we show the following results : 
\begin{eqnarray}
& &\bra{\psi}{\hat A}^{*2}\ket{\psi}
=\bra{\psi}{\hat A}^{*}\ket{\psi}^2[1-1/(2\Omega-4Y^*Y)]\nonumber\\
& &\qquad\qquad\qquad
+U^2V^{*2}(\bra{\psi}{\hat M}^{2}\ket{\psi}-\bra{\psi}{\hat M}\ket{\psi}^2)
/2\Omega \nonumber\\
& &\qquad\qquad\qquad
+U^4\bra{\psi}{\hat B}^{*2}\ket{\psi}
-2U^2V^{*2}\bra{\psi}{\hat B}^{*}{\hat B}\ket{\psi}
+V^{*4}\bra{\psi}{\hat B}^{2}\ket{\psi} \ , \quad
\label{4-34}\\
& &\bra{\psi}{\hat A}^{2}\ket{\psi}
=\bra{\psi}{\hat A}\ket{\psi}^2[1-1/(2\Omega-4Y^*Y)]\nonumber\\
& &\qquad\qquad\qquad
+U^2V^{2}(\bra{\psi}{\hat M}^{2}\ket{\psi}-\bra{\psi}{\hat M}\ket{\psi}^2)
/2\Omega \nonumber\\
& &\qquad\qquad\qquad
+V^4\bra{\psi}{\hat B}^{*2}\ket{\psi}
-2U^2V^{2}\bra{\psi}{\hat B}^{*}{\hat B}\ket{\psi}
+U^{4}\bra{\psi}{\hat B}^{2}\ket{\psi} \ , \quad
\label{4-35}\\
& &\bra{\psi}{\hat A}^{*}{\hat A}\ket{\psi}
=\bra{\psi}{\hat A}^{*}\ket{\psi}\bra{\psi}{\hat A}\ket{\psi}
+(X^*X)^2/(2\Omega(2\Omega-4Y^*Y))\nonumber\\
& &\qquad\qquad\qquad
+U^2V^{*}V(\bra{\psi}{\hat M}^{2}\ket{\psi}-\bra{\psi}{\hat M}\ket{\psi}^2)
/2\Omega-U^2V^2\bra{\psi}{\hat B}^{*2}\ket{\psi} \nonumber\\
& &\qquad\qquad\qquad
+(1-2U^2V^{*}V)\bra{\psi}{\hat B}^{*}{\hat B}\ket{\psi}
-U^2V^{*2}\bra{\psi}{\hat B}^{2}\ket{\psi} \ , \quad
\label{4-36}\\
& &\bra{\psi}{\hat A}{\hat A}^{*}\ket{\psi}
=\bra{\psi}{\hat A}^{*}{\hat A}\ket{\psi}
+(1-X^*X/\Omega-2Y^*Y/\Omega) \ , 
\label{4-37}\\
& &\bra{\psi}(1-{\hat N}/2\Omega)^{2}\ket{\psi}
=\bra{\psi}(1-{\hat N}/2\Omega)\ket{\psi}^2\nonumber\\
& &\qquad\qquad\qquad
+1/\Omega\cdot X^*X/\Omega\cdot(1-X^*X/2\Omega-4Y^*Y/2\Omega)/
(1-4Y^*Y/2\Omega)\nonumber\\
& &\qquad\qquad\qquad
+(1-2V^*V)^2\cdot1/4\Omega^2\cdot
(\bra{\psi}{\hat M}^{2}\ket{\psi}-\bra{\psi}{\hat M}\ket{\psi}^2)
\nonumber\\
& &\qquad\qquad\qquad
+2/\Omega\cdot U^2(V^2\bra{\psi}{\hat B}^{*2}\ket{\psi}
+2V^{*}V\bra{\psi}{\hat B}^{*}{\hat B}\ket{\psi}
+V^{*2}\bra{\psi}{\hat B}^{2}\ket{\psi}) \ . \nonumber\\
& &
\label{4-38}
\end{eqnarray}
Of course, the quantities such as $U$ which appear in the above 
expressions are replaced by Eqs.(\ref{4-20}) and 
(\ref{4-24})$\sim$(\ref{4-29}).

Now, we can discuss the uncertainty relation of our present system. 
For this purpose, it is necessary to show the expressions of 
$\Delta Q$ and $\Delta P$. 
With the use of the relations (\ref{4-35})$\sim$(\ref{4-38}) 
together with Eqs.(\ref{4-31}) and (\ref{4-32}), we can obtain the 
results shown as follows : 
\begin{eqnarray}
& &(\Delta Q)^2=\bra{\psi}{\hat Q}^2\ket{\psi}-\bra{\psi}{\hat Q}\ket{\psi}^2
\nonumber\\
& &\qquad\quad
=\hbar/2\cdot\biggl[\left(1-\frac{4Y^*Y}{2\Omega}\right)
\left(1-\frac{X^*X+X^{*2}}{2(\Omega-2Y^*Y)}\right)
\left(1-\frac{X^*X+X^2}{2(\Omega-2Y^*Y)}\right)\nonumber\\
& &\qquad\qquad\qquad\ \ 
+\frac{1}{2\Omega}U^2(V+V^*)^2
(\bra{\psi}{\hat M}^2\ket{\psi}-\bra{\psi}{\hat M}\ket{\psi}^2)\nonumber\\
& &\qquad\qquad\qquad\ \ 
+(U^2-V^2)^2\bra{\psi}{\hat B}^{*2}\ket{\psi}
+(U^2-V^{*2})^2\bra{\psi}{\hat B}^2\ket{\psi}\nonumber\\
& &\qquad\qquad\qquad\ \ 
+2(U^2-V^2)(U^2-V^{*2})\bra{\psi}{\hat B}^*{\hat B}\ket{\psi}\biggl] \ , 
\label{4-39}\\
& &(\Delta P)^2=\bra{\psi}{\hat P}^2\ket{\psi}-\bra{\psi}{\hat P}\ket{\psi}^2
\nonumber\\
& &\qquad\quad
=\hbar/2\cdot\biggl[\left(1-\frac{4Y^*Y}{2\Omega}\right)
\left(1-\frac{X^*X-X^{*2}}{2(\Omega-2Y^*Y)}\right)
\left(1-\frac{X^*X-X^2}{2(\Omega-2Y^*Y)}\right)\nonumber\\
& &\qquad\qquad\qquad\ \ 
-\frac{1}{2\Omega}U^2(V-V^*)^2
(\bra{\psi}{\hat M}^2\ket{\psi}-\bra{\psi}{\hat M}\ket{\psi}^2)\nonumber\\
& &\qquad\qquad\qquad\ \ 
-(U^2+V^2)^2\bra{\psi}{\hat B}^{*2}\ket{\psi}
-(U^2+V^{*2})^2\bra{\psi}{\hat B}^2\ket{\psi}\nonumber\\
& &\qquad\qquad\qquad\ \ 
+2(U^2+V^2)(U^2+V^{*2})\bra{\psi}{\hat B}^*{\hat B}\ket{\psi}\biggl] \ , 
\label{4-40}\\
& &\bra{\psi}{\hat Q}\ket{\psi}
=\sqrt{\frac{\hbar}{2}}(X^*+X)\sqrt{1-\frac{X^*X}{2\Omega}
-\frac{4Y^*Y}{2\Omega}} \ , 
\label{4-41}\\
& &\bra{\psi}{\hat P}\ket{\psi}
=i\sqrt{\frac{\hbar}{2}}(X^*-X)\sqrt{1-\frac{X^*X}{2\Omega}
-\frac{4Y^*Y}{2\Omega}} \ , 
\label{4-42}\\
& &\bra{\psi}{\hat R}\ket{\psi}
=\left(1-\frac{4Y^*Y}{2\Omega}\right)\left(1-\frac{X^*X}{\Omega-2Y^*Y}\right) 
\ .
\label{4-43}
\end{eqnarray}
With the use of the above relations, we can calculate 
$\langle ({\hat P}^2+{\hat Q}^2)/2 \rangle$, which is shown in the next 
section. 
The square of the quasi-spin is expressed as 
\begin{eqnarray}
& &\langle {\hat S}_x \rangle^2+\langle {\hat S}_y \rangle^2+
\langle {\hat S}_z \rangle^2=(\Omega-2Y^*Y)^2 \ , 
\label{4-44}\\
& &\langle {\hat S}_x^2 \rangle+\langle {\hat S}_y^2 \rangle+
\langle {\hat S}_z^2 \rangle=\Omega(\Omega+1) \ . 
\label{4-45}
\end{eqnarray}
Thus, the squeezed state gives the fluctuation of the components 
of the quasi-spin, which is represented by the variables $Y^*Y$.

\section{Expectation value for the Hamiltonian of the Lipkin model}

Let us consider the Hamiltonian (\ref{3-21}) in the Lipkin model. 
This Hamiltonian can be expressed in terms of the operators ${\hat N}$, 
${\hat A}$ and ${\hat A}^*$ as 
\begin{equation}\label{5-6}
{\hat H}=\frac{\epsilon}{2}\cdot{\hat N}-\frac{\chi}{2}
({\hat A}^{*2}+{\hat A}^2) -\epsilon\Omega \ . 
\end{equation}
The expectation values 
$\langle {\hat H} \rangle_{ch}$ and 
$\langle {\hat H} \rangle_{sq}$ with respect to both states 
$\ket{\phi}$ in (\ref{3-1}) 
and $\ket{\psi}$ in (\ref{4-1}), respectively, are easily evaluated by 
(\ref{3-8}), (\ref{3-9}), (\ref{4-33}), (\ref{4-34}) and (\ref{4-35}). 
In this section, we show the energy expectation values with respect to 
the state $\ket{\psi}$. 
Here, we introduce the other sets of canonical variables instead of 
$X$, $X^*$, $Y$ and $Y^*$, which 
correspond to the action and angle variables, as 
\begin{eqnarray}\label{5-2}
& &X=\sqrt{n_X}e^{-i\varphi_X} \ , \qquad X^*=\sqrt{n_X}e^{i\varphi_X} \ , 
\nonumber\\
& &Y=\sqrt{n_Y}e^{-i\varphi_Y} \ , \qquad Y^*=\sqrt{n_Y}e^{i\varphi_Y} \ . 
\end{eqnarray}
Since the state $\ket{\psi}$ has been constructed similar to the boson 
squeezed state, the variables $Y$ and $Y^*$ represent a certain kind of 
the fluctuation. 
Thus, $n_Y$ may be supposed to be a small value compared to the order 1. 
However, we calculate without an assumption of small fluctuations. 

We determine $\sqrt{n_Y}$ so as to guarantee the minimum uncertainty 
relation. Namely, we impose the condition that 
the introduced state $\ket{\psi}$ should retain to 
give the classical image. 
Under this condition, 
there are two ways to determine the $\sqrt{n_{Y}}$ to 
estimate the ground state energy.

\begin{figure}[t]
  \epsfxsize=7cm  
  \centerline{\epsfbox{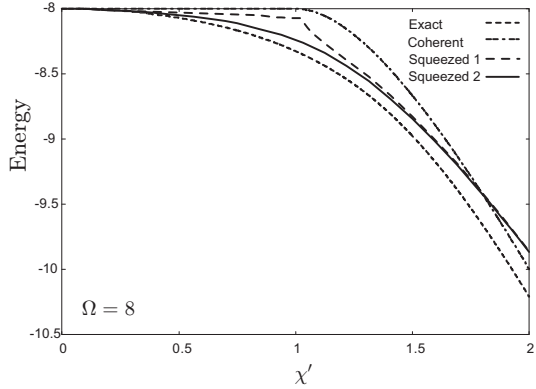}}
  \caption{Energy expectation values with respect to the squeezed state 
  $\ket{\psi}$ by using Method 1 (dashed curve), Method 2 (solid curve) 
  and the coherent state $\ket{\phi}$ (dash-dotted curve) are 
  depicted together with the exact eigenvalues (dotted curve) in the
  case $\Omega=8$. 
  The horizontal axis represents $\chi'$. }
  \label{fig:4}
\end{figure}

One way, which we call Method 1, is as follows: 
First, we set up phase factors as $\varphi_{X}=\varphi_{Y}=0$ because of 
the condition of energy minimum. 
Then, we determine the action variable $n_{X}$ from the condition 
$\partial \langle \hat{H}\rangle _{sq}/\partial n_{X}=0$.
After that, from the minimum uncertain relation 
\begin{eqnarray}
( \Delta \hat{Q}) ^{2}_{sq}
( \Delta \hat{P}) ^{2}_{sq}
=\frac{\hbar^{2}}{4}
|\langle \hat{R}\rangle_{sq}|^2 \ ,
\label{uncer}
\end{eqnarray}
we determine the $n_{Y}$.
Above-mentioned calculations should be carried out consistently.
By using these variables, the energy expectation value 
can be estimated. 

Another way, which we call Method 2, is as follows : 
First, we set up phase factor as $\varphi_{X}=\varphi_{Y}=0$
as well as the former way.
Secondly, we determine the $n_{Y}$ from the minimum uncertain relation
(\ref{uncer}).
After that, we seek the minimum energy expectation value 
with respect to $n_{X}$.
In this case, $n_{X}$ which satisfies
the condition 
$\partial \langle \hat{H}\rangle _{sq}/\partial n_{X}=0$
does not realize the energy minimum.

In Fig. 1, the energy expectation values obtained from 
Method 1 (dashed curve) and Method 2 (solid curve) 
are depicted compared with the 
exact ground state energy eigenvalues (dotted curve) 
with $\Omega=8$ and $\epsilon=1$. 
The horizontal axis represents $\chi'$ and the vertical axis represents 
the energy.
The expectation value with respect to $\ket{\psi}$ 
is close to the exact energy eigenvalue compared with the usual 
Slater determinant approach (dash-dotted curve). 
Especially, near the phase transition point, $\chi'=1$,
the energy expectation values
obtained by the 
squeezed state approach can trace the the exact energy eigenvalues 
approximately. 
It is pointed out, especially, that 
the energy expectation values near the transition point are well 
reproduced under the fixed minimum uncertainty relation (Method 2).

\section{Summary}

We have shown that an idea to extend the TDHF theory in the canonical 
form could be formulated based on the use of the state extended from the 
Slater determinant. 
The essential ingredients are to use 
the extended state from the Slater determinant and to impose 
the canonicity conditions. 
This extended state, which we call a quasi-spin 
squeezed state in the Lipkin model in our previous papers\cite{TKY94,TAKY96} 
was a kind of the squeezed state 
in comparison with the $su(2)$-coherent state. 
This state was constructed similar to the usual boson squeezed state. 
By imposing the canonicity conditions for the variables which characterizes 
the coherent part and the squeezed part, we could obtain 
the sets of canonical variables. 
Thus, it becomes possible to formulate the extended TDHF theory 
as a canonical form. 

As an application, the zero-point fluctuation induced by 
the uncertainty principle was investigated and the ground state 
energy was evaluated.
It has been shown that the ground state energy is well reproduced 
compared with the results obtained by using the Slater determinant. 
Especially, it was shown that, 
near the transition point, the energy expectation values 
calculated by imposing the condition of the fixed minimum uncertainty 
relation have reproduced well the exact energy eigenvalues. 

The dynamics will be investigated in our extended TDHF theory. 
This is a future problem.\cite{ATN03}

\acknowledgement 
The authors would like to express their sincere thanks to Professor 
M. Yamamura for giving them a chance to work about the extension of the 
TDHF theory in canonical form developed in this paper. 
This investigation is due in large part to his pioneering research.  
One of the authors (Y.T.) 
is partially supported by the Grants-in-Aid of the Scientific Research 
No. 15740156 from the Ministry of Education, Culture, Sports, Science and 
Technology in Japan.


\end{document}